\documentclass{aa}

\usepackage{graphicx}
\usepackage{txfonts}
\usepackage{color}

\begin{document}

\title{A Sagittarius-like simulated dwarf spheroidal galaxy from TNG50}

\author{Ewa L. {\L}okas
}

\institute{Nicolaus Copernicus Astronomical Center, Polish Academy of Sciences,
Bartycka 18, 00-716 Warsaw, Poland\\
\email{lokas@camk.edu.pl}}


\abstract{The Sagittarius dwarf spheroidal (Sgr dSph) galaxy provides one of the most convincing examples of tidal
interaction between satellite galaxies and the Milky Way (MW). The main body of the dwarf was recently demonstrated to
have an elongated, prolate, bar-like shape and to possess some internal rotation. Whether these features are temporary
results of the strong tidal interaction at the recent pericenter passage or are due to a disky progenitor is a matter
of debate. I present an analog of Sgr selected among bar-like galaxies from the TNG50 simulation of the IllustrisTNG
project. The simulated dwarf is initially a disky galaxy with mass exceeding $10^{11}$ M$_\odot$ and evolves around a
MW-like host on a tight orbit with seven pericenter passages and a period of about 1 Gyr. At the second pericenter
passage, the disk transforms into a bar and the bar-like shape of the stellar component is preserved until the end of
the evolution. The morphological transformation is accompanied by strong mass loss, leaving a dwarf with a final mass
of below $10^{9}$ M$_\odot$. The gas is lost completely and the star formation ceases at the third pericenter passage.
At the last pericenters, the dwarf possesses a bar-like shape, a little remnant rotation, and the metallicity gradient,
which are consistent with observations. The more concentrated metal-rich stellar population rotates faster and has a
lower velocity dispersion than the more extended metal-poor one. The
metallicity distribution evolves so that the most metal-poor stars are stripped first, which explains the metallicity
gradient detected in the Sgr stream. This study demonstrates that a dSph galaxy with properties akin to the Sgr dwarf
can form from a disky progenitor with a mass of above $10^{11}$ M$_\odot$ by tidal evolution around the MW in the
cosmological context.}

\keywords{galaxies: evolution -- galaxies: interactions --
galaxies: kinematics and dynamics -- galaxies: structure -- galaxies: dwarf -- Local Group}

\maketitle

\section{Introduction}

Discovered three decades ago \citep{Ibata1994}, the Sagittarius dwarf spheroidal (Sgr dSph) galaxy still provides the
most striking example of tidal interaction between the Milky Way (MW) and a satellite galaxy. Subsequent studies using
the data from the Two Micron All Sky Survey and the Sloan Digital Sky Survey (SDSS) led to detailed mapping of the core
of  Sgr and the tidal debris  \citep{Majewski2003, Newberg2002, Newberg2003, Newberg2007}. It has been demonstrated
that the main body of the dwarf is extremely elongated and that the tidally stripped stars form two long tidal tails
tracing an orbit that is almost perpendicular to the MW disk \citep{Martinez2004, Koposov2012, Belokurov2014,
Hernitschek2017, Antoja2020}. Kinematic measurements for Sgr member stars allowed observers to determine its velocity
dispersion profile \citep{Frinchaboy2012}, including its decrease toward the galaxy's center \citep{Majewski2013} and
revealed the presence of a significant rotation signal \citep{Lokas2010}.

More recently, \citet{delPino2021} performed a detailed study of the internal structure and kinematics of the core of
Sgr using machine-learning techniques, combining previous spectroscopic observations with new \textit{Gaia} data. These
authors derived the 3D positions and kinematics of more than $1.2 \times 10^5$ member stars, demonstrating that Sgr has
an almost perfectly prolate bar-like structure $\sim 2.5$ kpc long, with a short-to-long axis ratio of about 0.6. The
external regions of the main body were found to be expanding due to tidal forces, yet the galaxy conserves an inner
core of about 0.5 kpc that is not expanding and rotating with the velocity of about 4 km s$^{-1}$.

The dynamical status of the Sgr dwarf is still a matter of intense debate, with open questions concerning both the
origin of its elongated shape and the velocity gradient. It is not yet clear to what extent these features are of tidal
origin, resulting from the interaction with the MW or are signatures of the original properties of the progenitor dwarf
galaxy.

Many attempts have been made to reproduce the observed properties of Sgr using $N$-body simulations. Most of them used
a spherical progenitor \citep{Johnston1995, Ibata1997, Helmi2001, Vasiliev2020} and attempted to explain the present
elongated shape of the dwarf as resulting from the action of tidal forces at the last pericenter passage, which
probably occurred very recently. Only a few models were constructed with the Sgr progenitor in the form of a disk.
\citet{Penarrubia2010} proposed such a model in an attempt to reproduce the bifurcation in the Sgr stream. This model
was later rejected because it predicted too much rotation in the remnant, contrary to observations
\citep{Penarrubia2011}. The model of \citet{Lokas2010} avoided this problem by using a disk that produces a bar at the
first pericenter passage, which reduces the amount of rotation by introducing more radial orbits. A slightly modified
version of this model was found by \citet{delPino2021} to better reproduce the kinematics of Sgr measured with the
\textit{Gaia} data than the initially spherical models they also considered. However, exploring a wider family of
models, \citet{Vasiliev2020} claimed that similar data can still be reproduced with initially spherical progenitors.

In this paper, I try to shed some light on these issues using an analog of Sgr selected from the IllustrisTNG
simulations of galaxy formation \citep{Springel2018, Marinacci2018, Naiman2018, Nelson2018, Pillepich2018}. These
simulations solve for gravity and magnetohydrodynamics and apply additional prescriptions for AGN feedback,  star
formation, and galactic winds, allowing us to follow the evolution of dark matter and baryons in boxes of different
sizes and resolutions, and to study the properties of various objects, from dwarf galaxies to galaxy clusters. The
simulations were demonstrated to adequately reproduce many observed properties of galaxies, including their
morphologies \citep{Nelson2018, Genel2018, Rodriguez2019}. For the purpose of this study, I use the results of the
highest resolution TNG50 simulation \citep{Nelson2019b, Pillepich2019} performed in a box of about 50 Mpc in size,
which uses baryonic particles with masses of $8.5 \times 10^4$ M$_\odot$ and dark matter particles with masses of $4.5
\times 10^5$ M$_\odot$ with the corresponding softening scales of 0.074 and 0.288 kpc. The simulation data include 100
outputs, which allow us to follow the galaxy evolution with sufficient time resolution. The data are publicly available
and well documented, as described in \citet{Nelson2019a}.

The paper is organized as follows. In Section~2, I describe how the interacting system of interest was identified.
Section~3 presents the orbital evolution of the dwarf galaxy and Section~4 focuses on the properties of the tidally
induced bar. I discuss the measurements of the intrinsic and tidally induced rotation in the simulated dwarf in
Sect.~5, and the metallicity distribution and kinematics of different stellar populations in Sect.~6. The discussion
follows in Section~7.

\section{Identification of the interacting system}

The dwarf galaxy (with identification number ID117435 in the subhalo catalog of TNG50 at $z=0$) comes from the sample of
bar-like galaxies selected among almost 8000 galaxies of the last simulation output with stellar masses of $M_{*} >
3.4 \times 10^8$ M$_{\odot}$ within twice the stellar half-mass radius ($2 r_{1/2}$), and at least 100 stars, the
shapes of which were estimated by the Illustris team and provided in the Supplementary Data Catalogs of stellar
circularities, angular momenta, and axis ratios. The axis ratios were calculated as described in \citet{Genel2015} from
the eigenvalues of the stellar mass tensor obtained by aligning each galaxy with its principal axes and calculating
three components ($i$=1,2,3): $M_i = (\Sigma_j m_j r^2_{j,i}/\Sigma_j m_j)^{1/2}$, where $j$ enumerates over stellar
particles, $r_{j,i}$ is the distance of stellar particle $j$ in the $i$ -axis from the center of the galaxy, and $m_j$
is its mass. The eigenvalues were sorted so that $M_1 < M_2 < M_3,$ which means that the shortest to longest axis ratio
is $c/a = M_1/M_3$, while the intermediate to longest axis ratio is $b/a = M_2/M_3$.

\begin{figure}
\centering
\includegraphics[width=8cm]{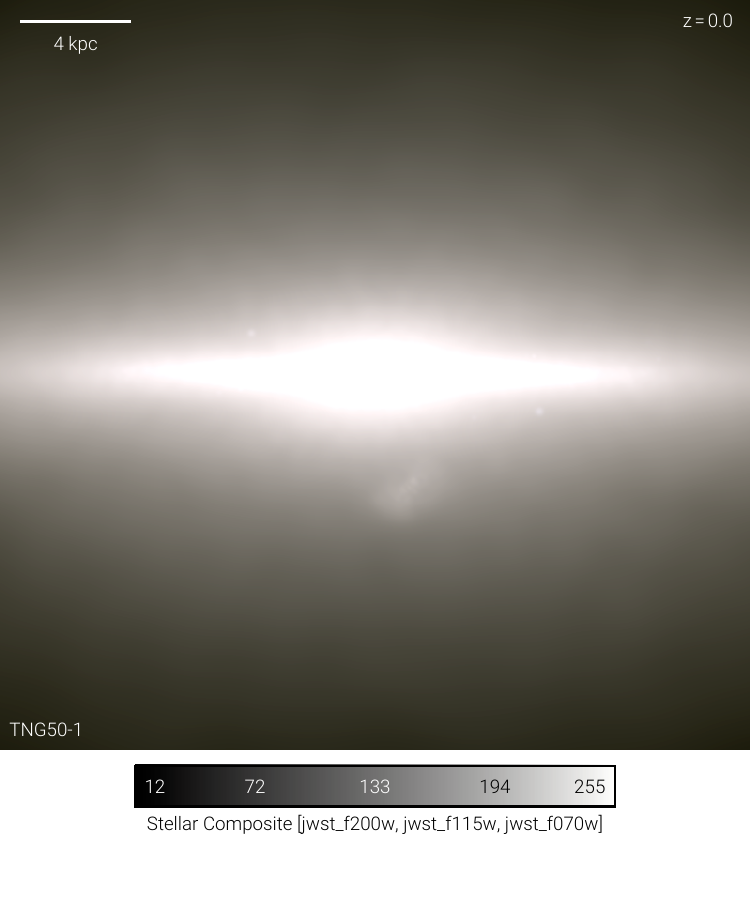}
\caption{Edge-on image of the host galaxy similar to the Milky Way (ID117252) with the dwarf remnant visible below the
disk.}
\label{image}
\end{figure}

The bar-like galaxies were then found by applying a single condition of the axis ratio $b/a < 0.6$ \citep{Lokas2021}.
This requirement yields a sample of 125 objects; however, after tracing their shape and mass evolution in time, I found
that a significant fraction of them are essentially dark-matter-free artifacts rather than real galaxy-hosting
subhalos. The mass and shape evolution of the selected dwarf stands out among the candidate objects, showing sudden
changes, which suggests a strong and persistent interaction with a bigger galaxy. A closer inspection of the dwarf's
history reveals that it indeed interacted with a galaxy similar in size to the MW; it survived a few pericenter
passages around it, and at the present time has just passed very close to the host. A composite image in JWST filters
of the host galaxy viewed edge-on  is shown in Fig.~\ref{image}, with the dwarf remnant visible below the disk.

The host galaxy (ID117252) is a massive spiral with a present mass of $1.3 \times 10^{12}$ and  $1.9 \times 10^{11}$
M$_{\odot}$ in the dark and stellar component, respectively. Around $t =11.7$ Gyr, it interacted with a cluster-size
object with a mass of above $10^{13}$M$_{\odot}$ with the closest approach being about 200 kpc. Before the interaction,
the dark mass of the galaxy reached $3.1 \times 10^{12}$M$_{\odot}$, while the gas mass was as high as $3.1 \times
10^{11}$M$_{\odot}$. During the interaction, the gas was strongly stripped so that at present its mass is only $3.2
\times 10^{9}$M$_{\odot}$. This interaction was one of the reasons the galaxy does not fulfill the isolation criteria
required to be included in the sample of MW/M31-like analogs previously selected from TNG50 \citep{Pillepich2023}.

\begin{figure}[!ht]
\centering
\includegraphics[width=8.4cm]{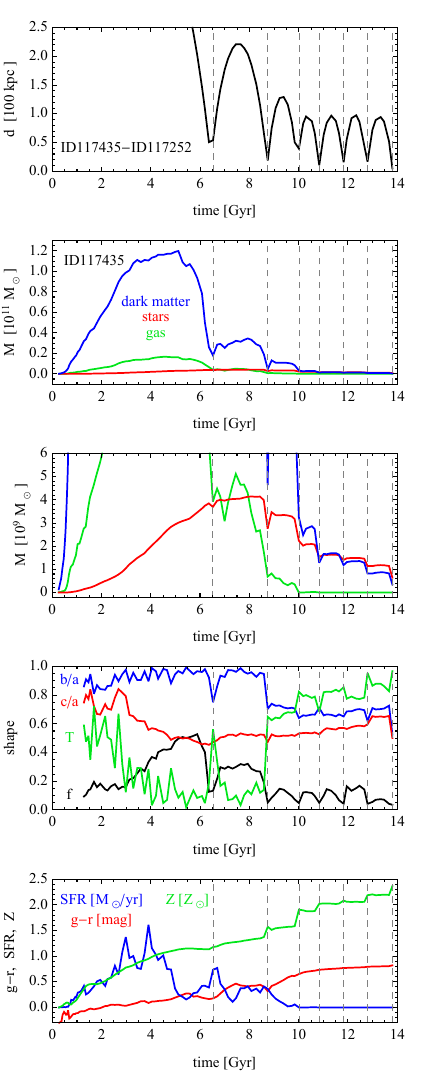}
\caption{Evolution of the Sgr-like dwarf galaxy ID117435. First panel: Distance of the dwarf from the host
(ID117252) as a function of time. Second and third panels: Mass of the dwarf in different mass scales. Fourth
panel: Evolution of the shape parameters (the axis ratios $b/a$, $c/a$ and triaxiality $T$) and the rotation parameter
$f$. Fifth panel: Evolution of the $g-r$ color, SFR, and metallicity $Z$. The vertical dashed lines in each panel mark the
pericenter passages.}
\label{evolution}
\end{figure}

\section{Orbital evolution of the dwarf galaxy}

The evolutionary history of the dwarf galaxy around the MW-like host is illustrated in Fig.~\ref{evolution}. The upper
panel shows the distance between the two galaxies as a function of time. The minima of the curve correspond to the
pericenter passages of the dwarf around the host and the maxima to the apocenters. The dwarf survives seven pericenters
in total. The first pericenter occurs around $t = 6.5$ Gyr and the subsequent ones at $t = 8.7, 10.0, 10.8, 11.8,
12.8,$ and 13.8 Gyr (vertical dashed lines in all panels of Fig.~\ref{evolution}). The last pericenter occurs close to
the present time, which for the cosmological model applied in the IllustrisTNG project corresponds to the age of the
Universe $t = 13.8$ Gyr. Due to the limited time resolution of the outputs available for study, the accuracy of
these pericenter times is about 0.1 Gyr. However, a closer inspection of the positions and velocities of the dwarf in
the last outputs reveals that at the end of the simulation the dwarf has just passed the last pericenter and is now at
about 5 kpc from the host. The radial orbital period between the last pericenters is around 1 Gyr. The pericentric
distances of the last few passages (before the final one) are around 16 kpc, while the apocenters occur around 96 kpc.
Thus, the pericentric values turn out to be relatively similar to what has been inferred for the real Sgr dwarf, while
the apocenters and therefore also the orbital period are higher \citep{Lokas2010, delPino2021}. It should be
emphasized, however, that the pericenter distance is more important for the evolution of the dwarf, because it is this
distance that determines how strongly the structure of the dwarf is affected by tidal forces from the host.

\begin{figure}
\centering
\includegraphics[width=8cm]{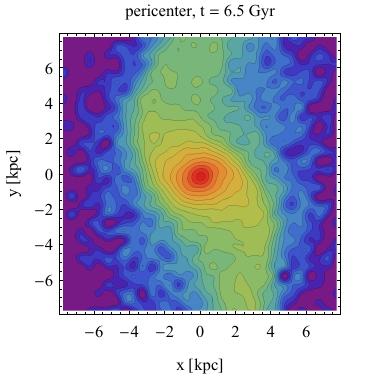}
\caption{Surface density distribution of the stellar component of the Sgr-like dwarf galaxy ID117435 viewed face-on
right after the first pericenter passage on its orbit around the host.}
\label{surdenfirst}
\end{figure}

The second and third panels of Fig.~\ref{evolution} show the evolution of the total mass of the dwarf in three
components (dark matter, stars, and gas) and two different mass scales to better illustrate the evolution at different
times. Before the interaction, the mass of the dark-matter halo of the dwarf reaches $1.2 \times 10^{11}$M$_{\odot}$, a
relatively large value, and characteristic of a normal-sized rather than a dwarf galaxy. At this point, the maximum
stellar and gas masses are $4.1 \times 10^{9}$M$_{\odot}$ and $1.7 \times 10^{10}$M$_{\odot}$, respectively. At the
first pericenter, the dwarf loses a significant part of its dark matter and gas due to tidal and ram-pressure
stripping, and continues to do so. The gas is lost almost completely at the third pericenter passage, around $t = 10$
Gyr. At the final simulation output ($t=13.8$ Gyr), the dwarf is reduced to a very small mass of $6.4 \times
10^{8}$M$_{\odot}$ in stars and $3.5 \times 10^{8}$M$_{\odot}$ in dark matter; that is, the amount of dark matter bound
to the dwarf is about twice smaller than that of the stars. It should be highlighted that at this stage the dwarf has
just passed a very tight pericenter of just a few kiloparsecs and is heavily truncated.

\begin{figure*}
\centering
\includegraphics[width=4.5cm]{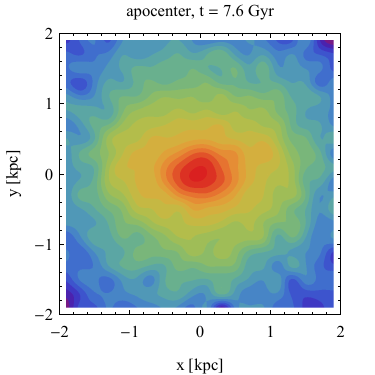}
\includegraphics[width=4.5cm]{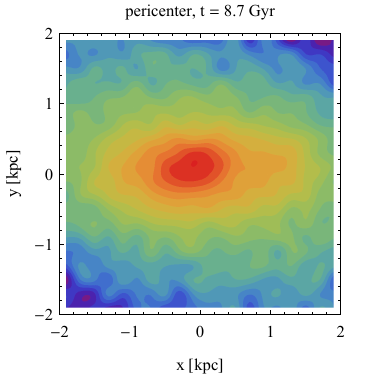}
\includegraphics[width=4.5cm]{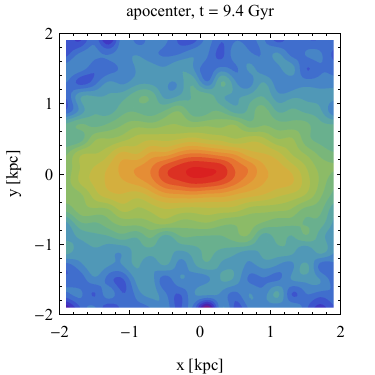}
\includegraphics[width=4.5cm]{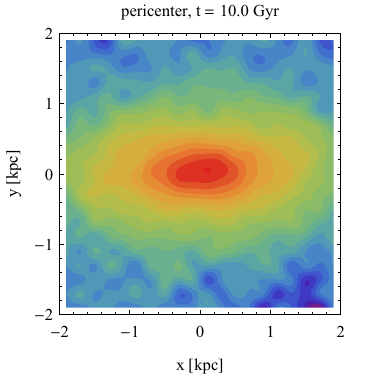}
\includegraphics[width=4.5cm]{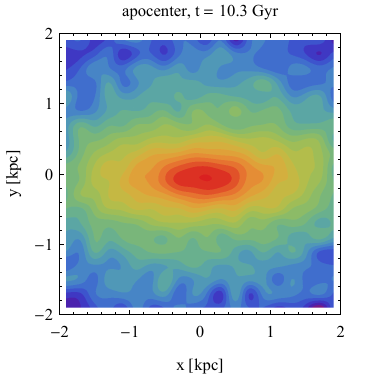}
\includegraphics[width=4.5cm]{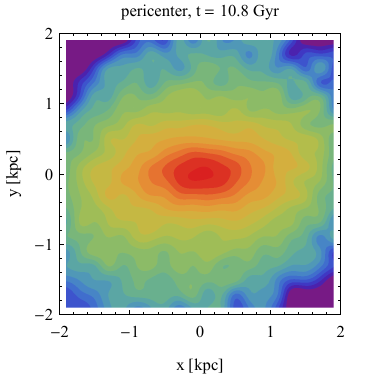}
\includegraphics[width=4.5cm]{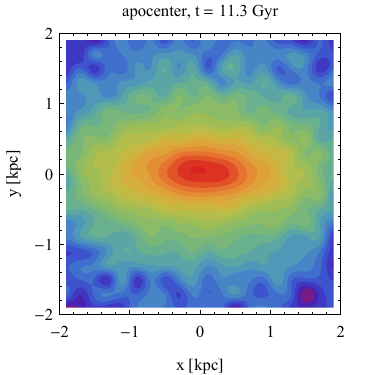}
\includegraphics[width=4.5cm]{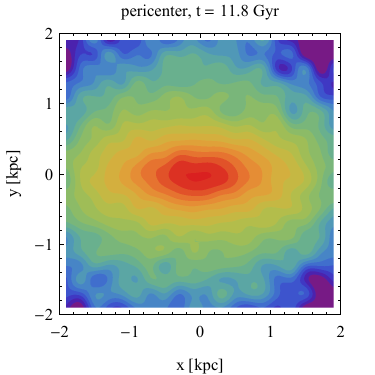}
\includegraphics[width=4.5cm]{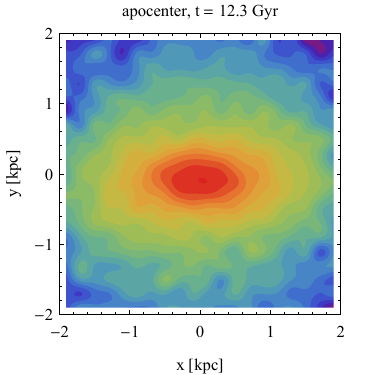}
\includegraphics[width=4.5cm]{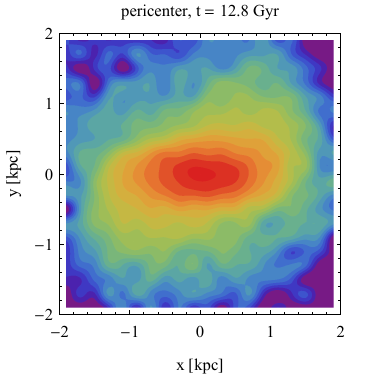}
\includegraphics[width=4.5cm]{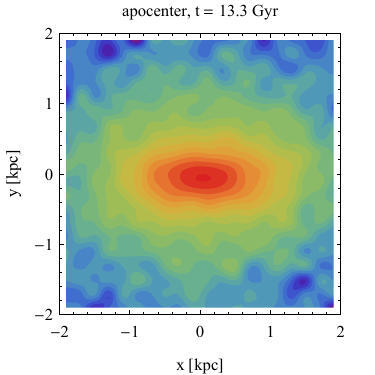}
\includegraphics[width=4.5cm]{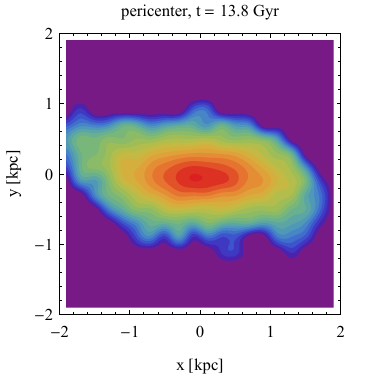}
\caption{Surface density distribution of the stellar component of the Sgr-like dwarf galaxy ID117435 viewed face-on at
different times. Surface density, $\Sigma,$ is normalized to the central maximum value in each plot and the contours
are equally spaced in $\log \Sigma$.}
\label{surden}
\end{figure*}

The fourth panel of Fig.~\ref{evolution} illustrates the evolution of different measures of the shape of the stellar
component: the axis ratios $b/a$ and $c/a$ as well as the triaxiality parameter $T = [1-(b/a)^2]/[1-(c/a)^2]$, all
calculated using stars within $2 r_{1/2}$. The triaxiality parameter provides a useful single-value measure of the
shape of the galaxy: it is low ($<1/3$) for oblate and high ($>2/3$) for prolate objects. One can see that until the
first pericenter passage around the host, the value of $b/a$ remains close to unity, while $c/a$ is significantly
lower, which means that the stellar component is in the shape of a disk at this stage. This is confirmed by the low
values of $T$. At the first pericenter passage, the disk is temporarily elongated ($b/a$ decreases while $T$
increases), but recovers the oblate shape soon after. The face-on view of the distorted disk right after the first
pericenter is shown in Fig.~\ref{surdenfirst}. It shows pronounced tidal arms that typically form in prograde
interactions of this kind; such interactions have been extensively studied in the literature \citep{Toomre1972,
Springel1999, Klimentowski2009, Lokas2015}.

The most interesting event in the dwarf's history takes place at the second pericenter passage, at $t = 8.7$ Gyr. At
that time, $b/a$ again decreases while $T$ increases dramatically, but in this case they retain their modified levels.
The value of $b/a$ stays at about 0.7, while $T$ varies slightly at subsequent pericenters but with an increasing
trend. This signifies the transition from the oblate to the prolate shape, that is the formation of the bar, which
later dominates the shape of the stellar component within $2 r_{1/2}$. The evolution of the bar is discussed further in
the following section.

The black line in the fourth panel of Fig.~\ref{evolution} additionally shows the rotation parameter, $f$, defined as
the fractional mass of all stars with circularity parameter $\epsilon > 0.7$, where $\epsilon=J_z/J(E)$, and $J_z$ is
the specific angular momentum of the star along the angular momentum of the galaxy, while $J(E)$ is the maximum angular
momentum of the stellar particles at positions between 50 before and 50 after the particle in question in a list where
the stellar particles are sorted by their binding energy \citep{Genel2015}. This parameter is a reliable measure of the
amount of rotation in the galaxy, with $f > 0.4$ considered to be characteristic of disks \citep{Joshi2020}. One can
see that prior to the first pericenter, the dwarf is indeed rotating in a way characteristic of disks, but the amount
of rotation decreases later on, until almost none is left at the end of evolution. This systematic decrease in the
rotational support is due to the transition from circular to more radial orbits of the stars in the bar.

The last panel of Fig.~\ref{evolution} illustrates the evolution of the properties of the stellar population of the
dwarf galaxy in terms of its color, star formation rate (SFR), and metallicity. The color is measured as the $g-r$
parameter (in terms of SDSS-like filters) with magnitudes based on the sum of the luminosities of all the stellar
particles of the subhalo. Because $g-r=0.6$ may be considered as a value dividing the red from the blue population
\citep{Nelson2018}, one can see (red line) that the dwarf evolves from the blue to the red object, passing the
threshold of $g-r=0.6$ at $t = 9.7$ Gyr, and ending up with $g-r=0.82$. In line with the color evolution, the SFR (blue
line) is largest in the early stages, when the dwarf is still a disk, and increases slightly during the first and
second pericenter passage as the gas is compressed by tidal forces. At the third pericenter, around $t = 10$ Gyr, the
SFR stops as the dwarf is completely depleted of gas. Finally, the green line plots the evolution of the metallicity of
the dwarf in terms of the mass-weighted average metallicity of all elements above He for stars within $2 r_{1/2}$.
Interestingly, the metallicity increases, showing variability controlled by the pericenter passages. This may indicate
that the different stellar populations are stripped differently and I discuss this issue in more detail in Section~6.

\section{Formation and evolution of the bar}

In the previous section, I describe the evolution of the shape of the stellar component of the dwarf, demonstrating the
transition from the oblate to the prolate shape, or in other words, the formation of the bar. The bar does not occupy
the very center of the disk, but rather almost the whole disk is transformed into a bar-like shape, similarly to
tidally induced bars in dwarf galaxies orbiting the MW and previously studied in controlled simulations
\citep{Lokas2014, Lokas2015, Gajda2017, Gajda2018}. Interestingly, as described in the previous section, in the case
studied here, the bar does not form after the first pericenter passage, despite the fact that the distortion of the
disk is very pronounced in Fig.~\ref{surdenfirst}. Let me note, however, that the inner part of the dwarf in
Fig.~\ref{surdenfirst} is not distorted and the surface density contours in this region remain circular. One may still
wonder why the permanent change into a bar-like shape happens at the second pericenter and not the first.  The effect
of tidal evolution depends first of all on the strength of the tidal force, which is larger for smaller pericenters.
When comparing the pericenters of the first two passages of the dwarf around the host, one can immediately see that
their distances are very different, around $d = 50$ kpc and $d = 19$ kpc, respectively, for the first and the second.

The effect of tides also depends on the inclination of the dwarf's disk  to the orbital plane. It has been demonstrated
\citep{Lokas2015, Lokas2018} that prograde encounters are more effective in inducing bars because of the resonant
behavior of the system. As explained by \cite{Lokas2018}, in the prograde configuration, the stars of the satellite are
affected by the tidal force for a longer time, increasing the likelihood of them being stripped. In the simulations
described here, the angle between the intrinsic and orbital angular momentum of the dwarf is 27 deg and 38 deg at the
first and the second pericenter, respectively. This means that although both are initially prograde, at the first
pericenter the configuration is more favorable to bar formation. However, it seems that the strength of the tidal
force --- which is greater at the second, tighter pericenter --- turns out to be more important. Another factor that may
play a role is the fact that, at the second pericenter, the dwarf is already significantly stripped and thus less
massive, and contains less gas, which makes it more susceptible to tidal transformation.

Figure~\ref{surden} shows the surface density distribution of the stellar component of the dwarf in the face-on
projection at different times, corresponding to subsequent apocenters and pericenters, from the apocenter at $t = 7.6$
Gyr in the upper left panel to the last pericenter at $t = 13.8$ Gyr in the lower right panel. The first image shows
the dwarf after the first pericenter passage, which creates only a temporary distortion and does not form a bar.
Indeed, one can see that in this image the dwarf recovers its disky shape and no bar is visible. At the second
pericenter, at $t = 8.7$ Gyr, the dwarf is already slightly distorted and forms a bar that is well visible in all the
remaining panels. It should be emphasized that the images at pericenters and apocenters are remarkably similar: the bar
is preserved during the whole evolution and the elongated shape is not due to the temporary distortion at pericenters.

\begin{figure}
\centering
\includegraphics[width=8cm]{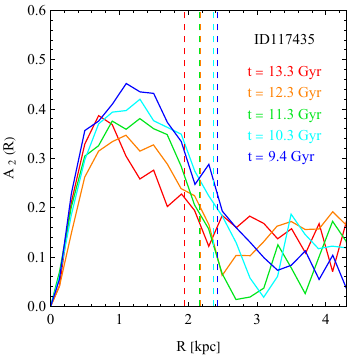}
\caption{Profiles of bar mode, $A_2 (R),$ for the Sgr-like dwarf galaxy ID117435 at the last five apocenters of its
orbit around ID117252. Measurements were carried out in bins of $\Delta R = 0.2$ kpc. Vertical dashed lines
indicate the corresponding bar lengths.}
\label{a2profiles}
\end{figure}

\begin{figure}
\centering
\includegraphics[width=9.cm]{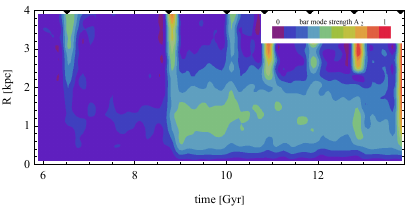}
\caption{Evolution of the profiles of the bar mode, $A_2 (R),$ of the Sgr-like dwarf galaxy ID117435 over time. The
region with the legend in the upper right corner, corresponding to later times and larger radii, is masked
because no measurements are available there. Black triangles near the upper horizontal axis indicate
pericenter passages.}
\label{a2modestime}
\end{figure}

\begin{figure*}
\centering
\includegraphics[width=4.5cm]{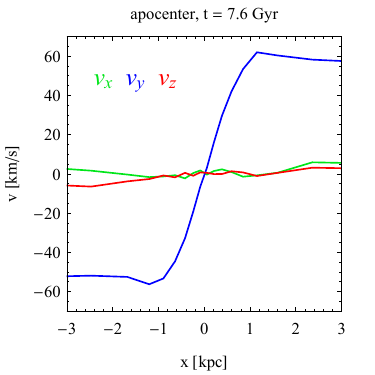}
\includegraphics[width=4.5cm]{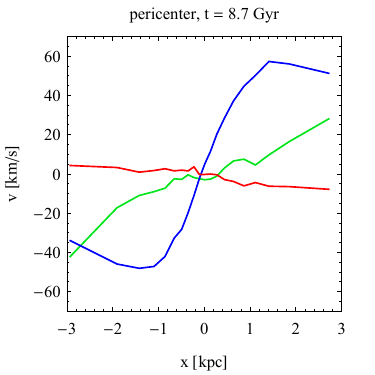}
\includegraphics[width=4.5cm]{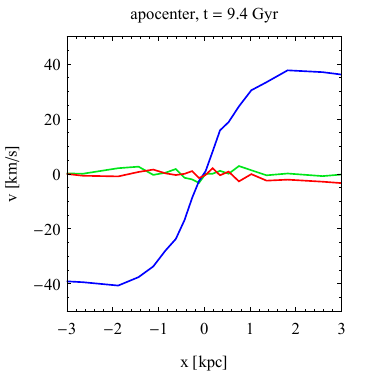}
\includegraphics[width=4.5cm]{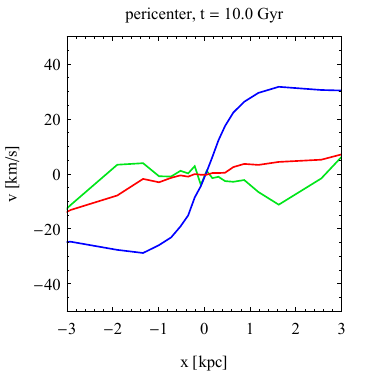}
\includegraphics[width=4.5cm]{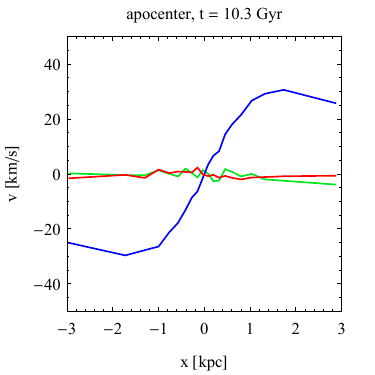}
\includegraphics[width=4.5cm]{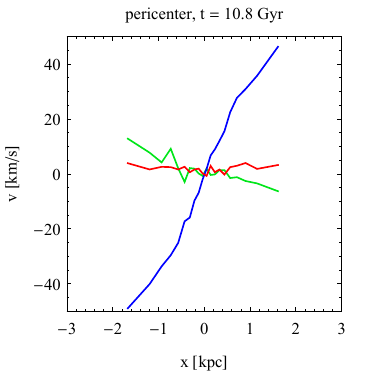}
\includegraphics[width=4.5cm]{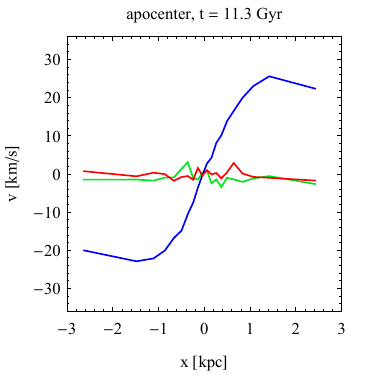}
\includegraphics[width=4.5cm]{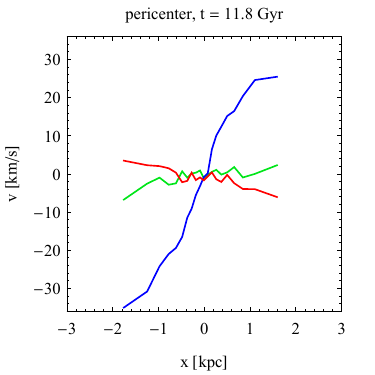}
\includegraphics[width=4.5cm]{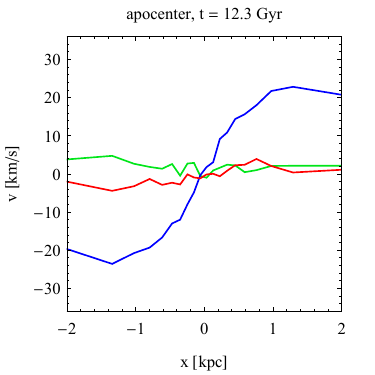}
\includegraphics[width=4.5cm]{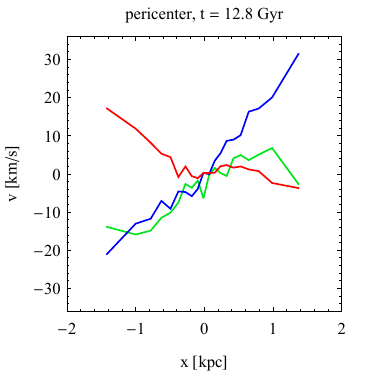}
\includegraphics[width=4.5cm]{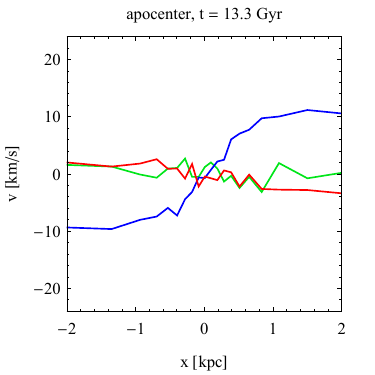}
\includegraphics[width=4.5cm]{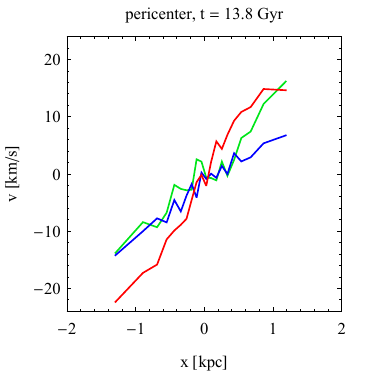}
\caption{Mean velocity of stars along the principal axes of the stellar component of the Sgr-like dwarf galaxy
ID117435, $v_x$, $v_y,$ and $v_z$, measured in 20 bins along the major axis $x$ at different times.}
\label{vel}
\end{figure*}

The evolution of the bar is best quantified using the commonly used measure of the strength of the bar
\citep{Athanassoula1986, Athanassoula2002, Athanassoula2013} in the form of the $m=2$ mode of the Fourier decomposition
of the surface density distribution of stellar particles projected along the short axis. This is given by $A_2 (R) = |
\Sigma_j m_j \exp(2 i \theta_j) |/\Sigma_j m_j$, where $\theta_j$ is the azimuthal angle of the $j$th star, $m_j$ is
its mass, and the sum goes up to the number of particles in a given radial bin. Figure~\ref{a2profiles} shows a few
examples of the $A_2 (R)$ profile at five subsequent apocenters when the bar is already formed and the dwarf is not
strongly affected by tides. The profiles display a shape characteristic of bars: an increase with radius, a maximum,
and a decrease. The bar is strongest at the apocenter of $t = 9.4$ Gyr, which is immediately after it forms at the
second pericenter passage. It then becomes weaker at subsequent apocenters due to tidal stirring, except for the last
apocenter, when it becomes stronger (but shorter) again. The length of the bar can be estimated as the radius $R$ where
the $A_2 (R)$ profile falls down to half the maximum value, $A_{2,{\rm max}}$. I calculated its values for the five
apocenters by fitting a straight line to the declining part of the $A_2 (R)$ profile and finding this radius. The
values of the bar length are indicated with vertical dashed lines in Fig.~\ref{a2profiles} and are found to decrease
almost monotonically with time, from 2.43 kpc at $t = 9.4$ Gyr to 1.95 kpc for $t = 13.3$ Gyr. This means that the bar
length is to some extent affected by the tidal forces at later stages, but remains close to the value of 2 kpc at all
times.

The full history of the bar can be appreciated by combining the bar mode profiles for different times into $A_2 (R,
t)$. This is shown in color-coded form in Fig.~\ref{a2modestime} for times later than $t = 6$ Gyr and up to the radius
of $R = 4$ kpc. As discussed above, at the time of the first pericenter passage, $t = 6.5$ Gyr, the dwarf galaxy was
strongly distorted, which is visible as a temporary increase in $A_2$ in Fig.~\ref{a2modestime}. However, the signal is
strong only at radii $R > 2$ kpc, while in the inner part the distortion is very weak, with $A_{2,{\rm max}} \sim 0.1$
and although it persists for about 2 Gyr, it does not lead to the formation of a bar. Instead, at the second, much
tighter pericenter passage, at $t = 8.7$ Gyr, a substantial distortion occurs also in the center of the dwarf and a
strong bar forms with $A_{2,{\rm max}} > 0.4$. The bar is permanent and survives until the end of the simulation, with
only small variations in strength and length. For example, it becomes stronger at the penultimate pericenter of $t =
12.8$ Gyr, but also shorter.

The consistently elevated values of $A_2$ within $R < 2$ kpc for all times in the range $9 < t < 13.8$ Gyr indicate the
presence of the bar, while the temporary increases at $R > 2$ kpc are due to the additional distortions during
pericenters. For the last four of them, the dwarf is so strongly truncated that no bound stars are present beyond $\sim
3$ kpc and no measurements of $A_2$ can be performed. This region has been masked with the legend in
Fig.~\ref{a2modestime}. At the later stages, the evolution of the bar is controlled, like its formation, by the
pericentric distance and the angle between the intrinsic and orbital angular momentum. In particular, the latter
orientation becomes less prograde during the last few gigayears of evolution. However, another effect comes into play
when the bar is already present, which is related to the orientation of the bar with respect to the tidal torque from
the host. As explained by \citet{Lokas2014}, this can lead to acceleration or deceleration of the bar, which translates
to its strength: slowing down makes the bar stronger while speeding up makes it weaker. The time resolution of the
available outputs of the TNG50 simulation ($0.1-0.2$ Gyr for the later snapshots) is not sufficient to study this
effect in detail; this would require a resolution of a few times higher, depending on the pattern speed of the bar.
However, its presence can be inferred from the behavior of the triaxiality parameter $T$ in the fourth panel of
Fig.~\ref{evolution} (green line). This parameter shows strong variation at pericenters, that is, both upward and
downward, which is most probably the result of the orientation of the bar with respect to the tidal force at the
pericenter.

\section{Intrinsic and tidally induced rotation}

One of the key open questions concerning the origin of the Sgr dSph galaxy refers to whether it was accreted as a disky
galaxy or was already spheroidal when it became a satellite of the MW. In the former case, the present elongation of
Sgr could be explained as a remnant of a bar originating from the disk instability, while in the latter the elongation
would simply be due to a temporarily present distortion caused by the tidal forces from the MW. In the scenario
involving a disky progenitor, one would expect the dwarf to possess significant remnant rotation. In order to
discriminate between these scenarios, it is thus essential to study not only the shape but also the kinematics of the
stars.

Using the data from the \textit{Gaia} survey, \citet{delPino2021} detected a significant rotation signal in the outer
parts of the dwarf, but not in the center. The interpretation of this result is difficult, because such rotation could
be either the remnant of the intrinsic rotation or induced by the tidal effects, or both. However, by comparing the
observed signal to the results of controlled simulations,  \citet{delPino2021} found that the simulation with a rotating
progenitor provides a better fit to the data than the one with a nonrotating progenitor. Here, I try to provide some
insight into this issue by looking at the evolution of the rotation in the Sgr analog from TNG50.

Figure~\ref{vel} shows the mean velocity of stars bound to the simulated dwarf along the principal axes of the stellar
component $v_x$, $v_y,$ and $v_z$ (where $x$, $y,$ and $z$ are measured along the longest, intermediate, and shortest
axis, respectively). The stars were binned into 20 bins containing the same number of stars along the major axis $x$.
The outputs used in this figure are the same as those in Fig.~\ref{surden}, namely they correspond to the subsequent
apocenters and pericenters starting from the apocenter at $t = 7.6$ Gyr (when the dwarf is still a disk) and ending
with the pericenter at $t = 13.8$ Gyr. In these measurements, the rotation along the $y$ axis, $v_y$ (blue line),
corresponds to the intrinsic rotation of the disk that would be oriented counterclockwise in the face-on view of the
images of Fig.~\ref{surden}. In an isolated object in equilibrium, this should be the only significant rotation whether
it is a disk or a tumbling bar, or a combination of those.

As can be seen in Fig.~\ref{vel}, the rotation signal in $v_y$ (blue line) is similar in the apocenters and
pericenters, although some evolution can be seen. First of all, the overall rotation signal diminishes with time due to
tidal stirring --- which turns the ordered motion of the stars into random motions --- and mass loss (less rotational
support is needed in a less massive dwarf). There is also a significant modification of the shape of the rotation curve
when passing from the apocenter to the following pericenter (each pair of these plots has the same velocity scale range
to make the comparison easier). Namely, the curve is straighter at the pericenter, being extended by the tidal forces.
This is especially clearly seen in the two left images of the middle row, for the apocenter at $t = 10.3$ Gyr and the
pericenter at $t = 10.8$ Gyr.

The nonzero velocity values of the other components, $v_x$ and $v_z$, are entirely due to the tidal forces. Again, a
comparison between the apocenters and pericenters reveals that at the apocenters the values of $v_x$ and $v_z$ are much
smaller, as expected given the weaker tidal forces present at larger distances of the dwarf from the host. Instead, at
pericenters, significant velocity gradients are seen in these components, depending on the particular orientation of
the dwarf's bar with respect to the tidal force.

The velocity gradients are especially interesting in the final simulation output, at $t = 13.8$ Gyr, corresponding to
the present time. In this case, all the gradients are strong, but those of $v_x$ (green line) and especially $v_z$ (red
line) are even stronger than that of $v_y$ (blue line). In this particular configuration, the tidally induced velocity
gradients become larger than the remnant rotation of the original disk. It should be emphasized that this stage of the
evolution of the simulated dwarf cannot be treated as an exact analog of the real Sgr, because the simulated object is
then very close to the host, at $d \sim 5$ kpc, and has just passed even closer. The corresponding distance of the real
Sgr is much larger, of about 18 kpc. These measurements demonstrate, however, that in extreme conditions the tidally
induced rotation signal at the later stages of evolution can become comparable to the remnant rotation even if the
latter was initially quite strong.

\begin{figure}
\centering
\includegraphics[width=8cm]{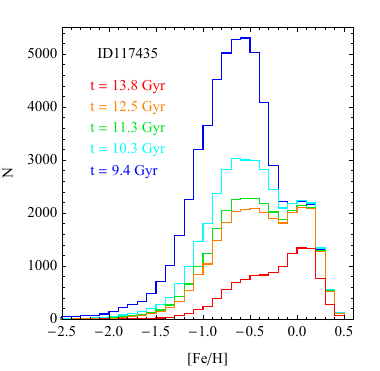}
\caption{Metallicity distribution of stars belonging to the Sgr-like dwarf galaxy ID117435 at the last four apocenters
and the last pericenter of its orbit around the host.}
\label{histograms}
\end{figure}

\begin{figure*}
\centering
\includegraphics[width=8cm]{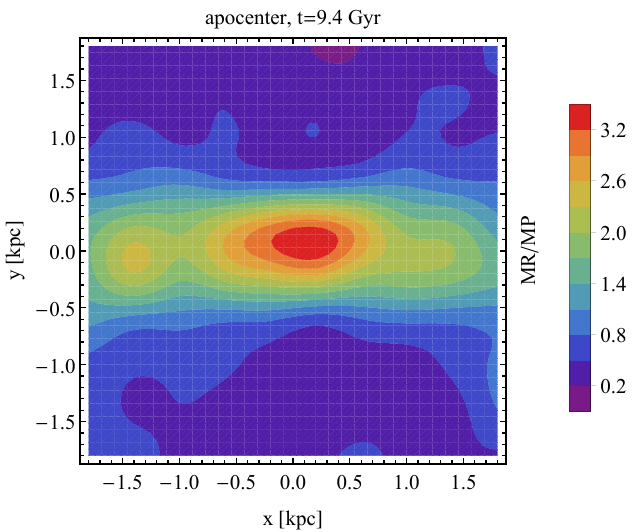}
\includegraphics[width=8cm]{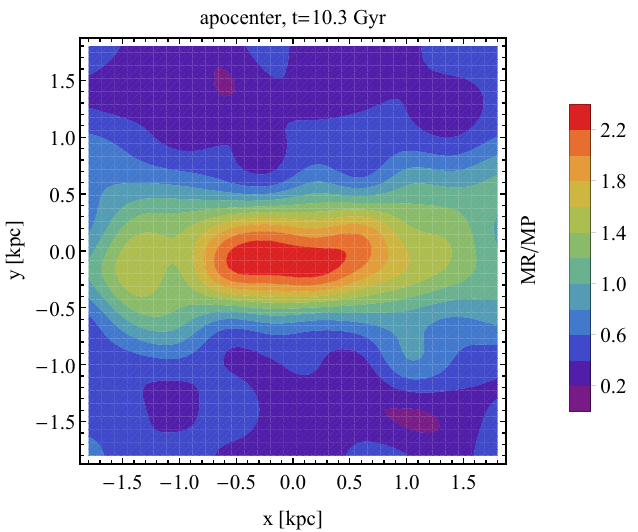}
\includegraphics[width=8cm]{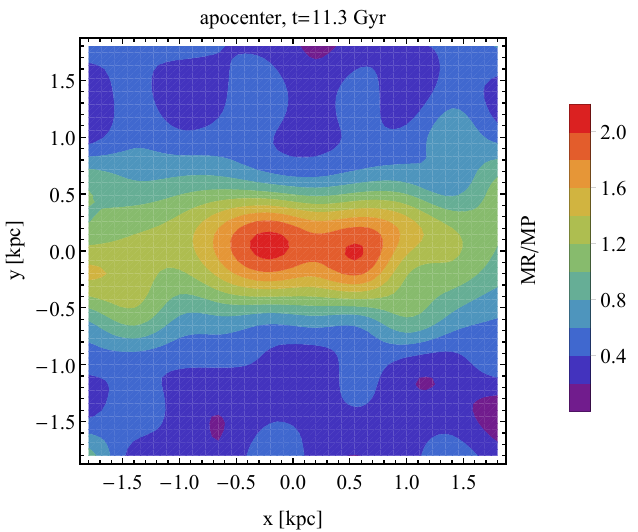}
\includegraphics[width=8cm]{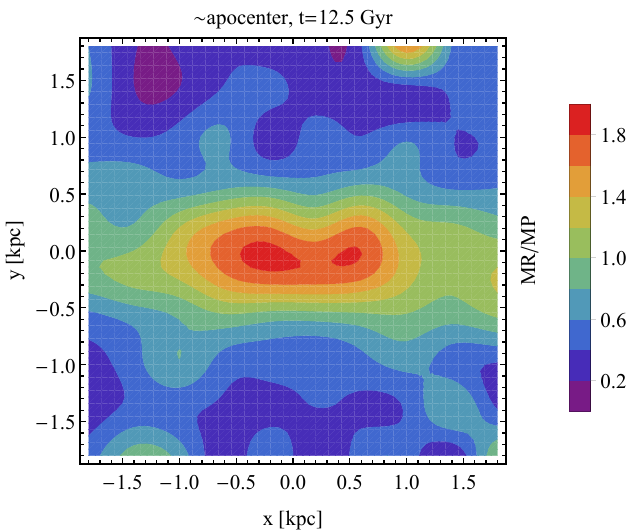}
\caption{Maps of the ratio of the number of metal-rich (MR) to metal-poor (MP) stars belonging to the Sgr-like dwarf
galaxy ID117435 at the last four apocenters of its orbit around the host. The measurements were taken in the face-on
projection in $10 \times 10$ cells of size 0.4 kpc.}
\label{mrmp}
\end{figure*}

\begin{figure}
\centering
\includegraphics[width=4.4cm]{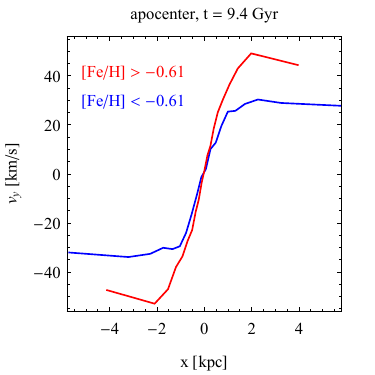}
\includegraphics[width=4.4cm]{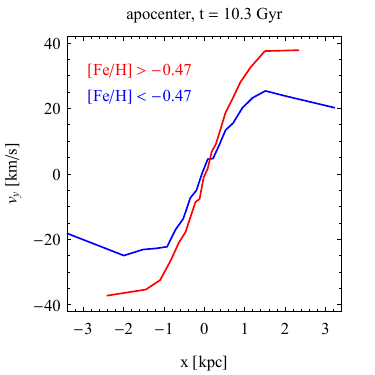}
\includegraphics[width=4.4cm]{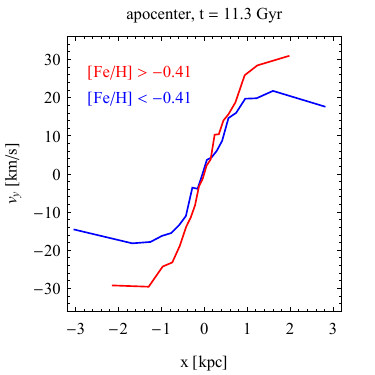}
\includegraphics[width=4.4cm]{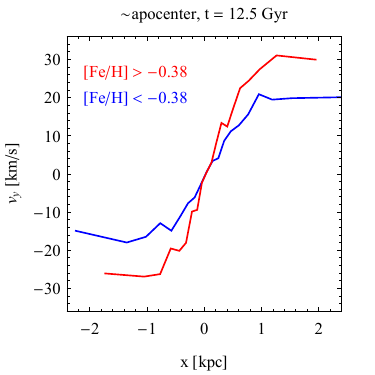}
\caption{Mean velocity along the intermediate axis $v_y$ measured in 20 bins along the major axis $x$ for stars of
different metallicity belonging to the Sgr-like dwarf galaxy ID117435 at the last four apocenters of its orbit around
the host. The stars were divided into two equal-size populations of metal-rich (red lines) and metal-poor (blue
lines) by metallicity threshold corresponding to the median value at each time.}
\label{velpop}
\end{figure}

\begin{figure}
\centering
\includegraphics[width=4.4cm]{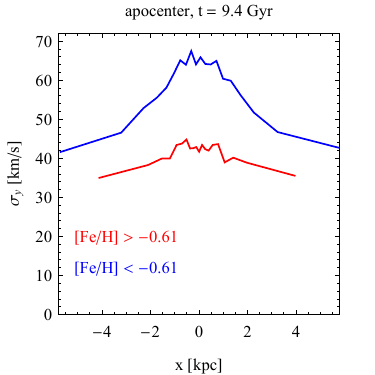}
\includegraphics[width=4.4cm]{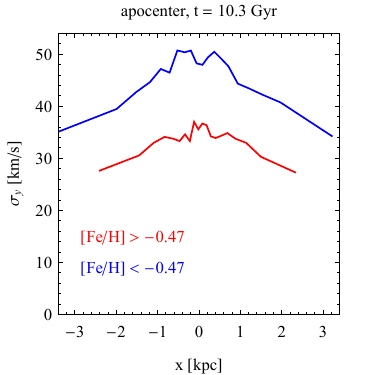}
\includegraphics[width=4.4cm]{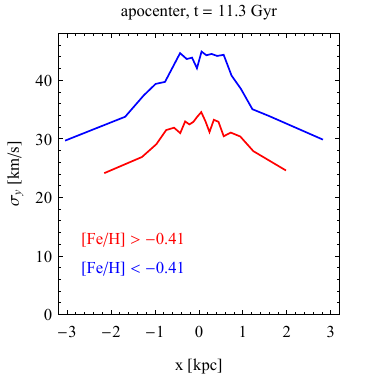}
\includegraphics[width=4.4cm]{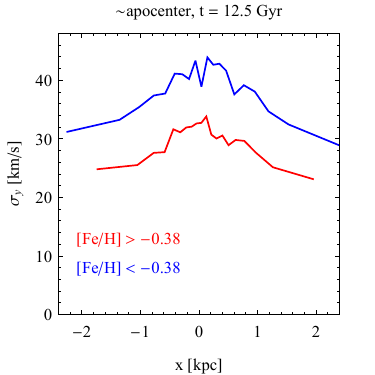}
\caption{Same as Fig.~\ref{velpop}, but for the velocity dispersion.}
\label{disppop}
\end{figure}

\section{Metallicity and kinematics}

Additional hints concerning the possible origin of the Sgr dwarf can be obtained using metallicity as well as combining
data on kinematics and metallicity. In section~2, I show (in the lower panel of Fig.~\ref{evolution}) that the total
metallicity of the simulated dwarf evolves strongly in time, increasing systematically with each pericenter passage.
These measurements correspond to the mass-weighted average metallicity of all elements above He in stars bound to the
dwarf galaxy at each output and contained within $2 r_{1/2}$. However, to facilitate the comparison with observations,
it is more useful to study the iron abundance [Fe/H]. The IllustrisTNG data include 20 full snapshots for which the
information on gas particles contains individual abundances of nine species including Fe in terms of the dimensionless
ratio of mass in that species to the total gas cell mass. For the stellar particles, the metals are inherited from the
gas cell converted into a given star, at the time of birth. This data can be expressed in terms of the number of nuclei
rather than mass and measured with respect to the solar abundance, as is usually done in observations.

Luckily, the 20 full snapshots for which these detailed data are available include five that are of particular interest
in the case of the simulated dwarf studied here. These five outputs (number 72, 78, 84, 91, and 99) correspond to four
last apocenters and the last pericenter of the dwarf's orbit around its host at $t = 9.4, 10.3, 11.3, 12.5,$ and 13.8
Gyr (only output 91 comes a little after the apocenter I discuss above, which occurs at output 90, $t = 12.3$ Gyr).
Figure~\ref{histograms} shows the iron abundance distribution in the dwarf for all stars bound to it at these five
times. One can see that at earlier times (especially at $t = 9.4$) the distribution is strongly peaked around $-0.5$
and is remarkably similar to the distribution for the real Sgr recently measured by \citet{Minelli2023} (their fig. 8),
except for a significant additional contribution from more metal-rich stars with $0 <$ [Fe/H] $<0.5$ in the simulation.
I discuss the possible origin of this difference in Sect. 7.

It is clear from Fig.~\ref{histograms} that as the dwarf galaxy evolved, at the four subsequent apocenters the
metal-poor population was diminished while the metal-rich one stayed almost intact. A departure from this behavior is
only seen in the last simulation output at $t = 13.8$ Gyr, where the dwarf is at the extremely close pericenter and
both populations are heavily stripped; however, even in this case the metal-poor population is affected to a greater
extent and the metal-rich stars start to dominate. This overall trend of preferential stripping of metal-poor stars
translates into increasing values of the median stellar iron abundance in the dwarf, which are $-0.61, -0.47, -0.41,
-0.38,$ and $-0.18$ for $t = 9.4, 10.3, 11.3, 12.5,$ and 13.8 Gyr, respectively.

To understand this evolution of metallicity, it is useful to study the spatial distribution of metal-poor and
metal-rich stars. To this end, I divided the stars belonging to the dwarf at each output into equally numerous
metal-poor and metal-rich population using the median [Fe/H] values given above. Figure~\ref{mrmp} shows the ratio of
the number of metal-rich to the number of the metal-poor stars in the face-on view at
four apocenters of $t = 9.4, 10.3, 11.3$, and $12.5$ Gyr in a way similar to \citet{Vitali2022} for the real Sgr. The
figure demonstrates that the metal-rich stars are more abundant in the center of the simulated dwarf at all stages of
evolution, although this difference is even greater at earlier times. This latter trend may seem to contradict the
conclusions drawn from Fig.~\ref{histograms}, which suggests that the metal-poor stars are more abundant at earlier
times, but it only reflects the fact that the definitions of the metal-poor and metal-rich populations are different at
different times because the median [Fe/H] values evolve. Adopting the same fixed metallicity threshold for all times to
define populations would reveal the opposite trend.

It turns out that the metal-poor and metal-rich populations defined with the median values not only have different
spatial distributions but also different kinematics. Figures~\ref{velpop} and \ref{disppop} show the mean velocity and
velocity dispersion along the $y$ axis of the simulated dwarf, $v_y$ and $\sigma_y$, separately for the metal-poor and
metal-rich populations at the four apocenters of $t = 9.4, 10.3, 11.3,$ and 12.5 Gyr. These measurements correspond to
the rotation velocity and velocity dispersion a distant observer would measure looking at the dwarf edge-on and binning
the data along the major axis, $x$. One can see that, in all cases, the more concentrated, metal-rich population (red
lines) rotates faster and has lower velocity dispersion values than the metal-poor population. This trend agrees very
well with the results for the real Sgr dwarf by \citet{Minelli2023} (their fig. 9), although their values for both the
rotation velocity and velocity dispersion are systematically lower.

\section{Discussion}

In this paper, I present an analog of the Sgr dSph galaxy identified among bar-like galaxies in the TNG50 simulation.
The simulated dwarf shares many similarities with the real Sgr, including the order of magnitude of mass, the elongated
shape, the remnant rotation, and the long history of past interaction with a MW-like host. This study demonstrates that
it is possible to produce a Sgr-like object from a massive, disky progenitor evolving on a tight orbit around its host
in the cosmological context, thus validating a previously proposed  scenario studied in controlled simulations
\citep{Lokas2010, delPino2021}. Interestingly, the maximum mass reached by the progenitor is as high as $1.4 \times
10^{11}$ M$_\odot$, which is on the order of the progenitor masses found to be consistent with the data if the Sgr
dwarf started its evolution on an orbit with an apocenter above$~200$ kpc \citep{Jiang2000, Gibbons2017}.

It is not clear, which stage (pericenter) provides the best match between the simulated dwarf and the real Sgr. At
the final snapshot of the simulation, the dwarf finds itself at a pericenter that is closer to the host (at $~\sim 5$
kpc) than the real Sgr. The end product of the simulation may therefore rather illustrate the future of the real Sgr,
and the present state of the real Sgr would correspond to one of the earlier pericenters of the orbit. This
interpretation is favored by the metallicity distribution (simulated final state is too metal-rich) and the time of gas
loss. The real Sgr contains no gas at present \citep{Koribalski1994, Burton1999}, but lost its gas more recently than
the simulated galaxy because it contains a population of relatively young stars of around 2 Gyr in age and near solar
metallicity, and a possibly even younger and more metal-rich component \citep{Siegel2007}. Taking into account all
properties, it seems that the best approximation of the real Sgr is provided by the simulated dwarf at $t = 12.8$ Gyr,
which corresponds to the penultimate pericenter. At this stage, the dwarf is at the right distance from the host ($d =
16.7$ kpc), the mass of the dwarf is already as low as $2 \times 10^9$ M$_\odot$, it has strongly diminished
intrinsic rotation, a decidedly prolate, elongated shape, and two well-defined populations of stars with different
metallicity and kinematics. However, the orientation of the intrinsic angular momentum with respect to the orbital one
is no longer prograde, which hampers any detailed comparison of kinematics with the real data.

The most important feature of the Sgr dwarf in the scenario with a disky progenitor is the remnant intrinsic rotation.
Such rotation is indeed present at all times in the simulated dwarf. However, as discussed in detail by
\citet{Martinez2023} for the population of MW/M31 satellites in TNG50, an additional rotation component comes into play
immediately after pericenters, originating from the tidal torque induced by the host. These rotation signals have
similar dependence on radius, namely they are both expected to be stronger further from the dwarf's center, and this is
indeed what is seen for the Sgr-like dwarf studied here. For this particular case, the two rotation signals become
comparable at the final simulation output when the dwarf is particularly strongly affected by tides. This means that
the rotation detected in Sgr using \textit{Gaia} data \citep{delPino2021}, which is increasing with radius, can be both
of intrinsic and tidal origin, since Sgr happens to have just passed its pericenter. Therefore, this rotation signal
does not allow us to discriminate between different scenarios or claim that Sgr indeed originated from a disk.

Additional constraints on the properties of the progenitor of the Sgr dwarf could be obtained from the metallicity
gradient and the kinematics of different stellar populations. However, no strong correlations have been found so far
between the metallicity gradient of the galaxies of the Local Group and other properties, such as the morphological
type, stellar mass, or the amount of rotation \citep{Taibi2022}. The fact that the Sgr dwarf still has a substantial
metallicity gradient today cannot therefore be used as an argument in favor of its origin as a disky galaxy. Still, the
simulated dwarf presented here does originate from a disk and indeed reproduces the metallicity and kinematic
properties observed in the real Sgr. In particular, I demonstrate that the metal-rich stars dominate in the center of
the dwarf over a long time in the later stages of evolution, as observed for the real Sgr by \citet{Majewski2013} and
\citet{Vitali2022}. The stellar populations of different metallicity in the simulated dwarf have different kinematic
properties, with the metal-rich stars rotating faster and having lower velocity dispersion than the metal-poor ones, in
agreement with the trends measured for the real Sgr by \citet{Majewski2013} and \citet{Minelli2023}.

The overall stellar metallicity distribution of the simulated dwarf in the intermediate stages is also similar to the
one published by \citet{Minelli2023} with a peak at [Fe/H]$=-0.5$ (their fig. 8) but has an additional, more metal-rich
component not present in their data. This may be due to the fact that the measurements of \citet{Minelli2023} were
performed in four regions not including the very center of Sgr, where the most metal-rich population should be present.
In fact, the bimodal metallicity distribution of the simulated dwarf in the later stages is quite similar to the one
obtained by \citet{Majewski2013}, where the central region was studied. Moreover, the evolution of the metallicity
distribution and the kinematic properties of different populations in the simulated dwarf are consistent with the
observed metallicity distribution and kinematics of the Sgr stream. In particular, \citet{Gibbons2017} found that the
stream contains a metal-rich, colder population and a metal-poor one with a higher velocity dispersion. Such a
composition arises naturally from stripping a dwarf galaxy containing stellar populations with similar properties. The
metallicity of the stars in the stellar stream varies from [Fe/H] $\sim -2.5$ to $\sim -0.5$ \citep{DeBoer2015} in good
agreement with the range of metallicities of stars stripped from the simulated dwarf galaxy studied here. Metallicity
gradients in the Sgr stream consistent with such a stripping scenario were recently detected using APOGEE
\citep{Hayes2020} and \textit{Gaia} data \citep{Limberg2023, Cunningham2024}.

The morphology of the stream containing stars tidally stripped from the simulated dwarf is not, however, very similar
to its real counterpart. Parts of the stream can be identified in the simulation by tracing stars that were bound to
the dwarf a few snapshots earlier and now belong to the host galaxy. These parts indeed form elongated tidal arms
typically produced in such interactions. Overall, however, the structure of the material lost from the dwarf is more
diffuse and complicated, because by the end of the simulation some of the stars have already fallen through the center
of the host, forming shells and dissipating. This is expected for the strongly nonspherical and evolving potential of
the host galaxy \citep{Ibata2001}, which, unlike the real MW, has itself interacted with a larger neighbor in its past.

\begin{acknowledgements}
I am grateful to the IllustrisTNG team for making their simulations publicly available and to the anonymous
referee for useful comments.
\end{acknowledgements}

\end{document}